# Spin wave surface states
# in one-dimensional planar magnonic crystals

**J Rychły, J W Kłos**

Faculty of Physics, Adam Mickiewicz University in Poznan, Umultowska 85, 61-614 Poznań, Poland

e-mail: rychly@amu.edu.pl

**Abstract**. We have investigated surface spin wave states in one-dimensional planar bi-component magnonic crystals, localized on the surfaces resulting from the breaking of the periodic structure. The two systems have been considered: the magnonic crystal with periodic changes of the anisotropy field in exchange regime and the magnonic crystal composed of Fe and Ni stripes in dipolar regime with exchange interactions included. We chose the symmetric unit cell for both systems to implement the symmetry related criteria for existence of the surface states. We investigated also the surface states induced by the presence of perturbation of the surface areas of the magnonic crystals. We showed, that the system with modulated anisotropy is a direct analog of the electronic crystal. Therefore, the surface states in both systems have the same properties. For surface states existing in magnonic crystals in dipolar regime we demonstrated that spin waves preserve distinct differences to the electronic crystals, which are due to long-range dynamic dipolar interactions. We found that tuning of the strength of magnetization pinning resulting from the surface anisotropy or dipolar effect is vitally important for existence of the surface states in magnonic crystals.

## 1. Introduction

Surfaces are the natural limitation of any real physical system. The physical processes are determined and described not only by the particular form of differential equations but also by the geometry of surfaces and the boundary conditions on the surfaces [1].

Surfaces play an important role in magnonics [2,3]. Constraints of the systems influence both the static magnetic configuration and the dynamic properties of the systems. For the magnetization dynamics both the static effective field describing the landscape perceived by all spin waves (SWs) (independent on magnitude and the direction of the wave vector), and the dynamical field determining the dynamical coupling of precessing magnetic moments are important. Even for the systems in the saturation state, where the magnetization is collinear and constant in magnitude, the (nonlocal) static demagnetizing field can appear as a result of the presence of surfaces and interfaces. Also, the magnetocrystalline anisotropy field can be induced (locally) by the presence of surfaces/interfaces in Ref.4. Both fields determine also the strength of SW pinning on the surfaces. The dynamical demagnetizing field also can contribute to SW pinning on the surface [5,6]. The surface anisotropy introduces additional torque acting on magnetic moments on the surface and can lead to the SW pinning or additional freedom depending on its orientation [7,8]. The surface/interface induced dynamical magnetization field is also crucial for ensuring the coupling between precessing magnetic moments which is necessary for SW propagation with non-zero group velocity dependent on the direction and magnitude of wave vector [9,10]. Summarizing, the surfaces are responsible for the following features of magnonic system: (i) formation of static magnetic configuration for given external magnetic field due to spatial constrains and induction of the static demagnetizing field [10,11], (ii) forcing boundary condition on the magnetization dynamics at the surfaces (external interfaces between magnetic and nonmagnetic material) due to surface anisotropy or dipolar pinning [12,8] (iii) coupling of the precession of magnetic moments and tuning the group velocity of SW precession due to dynamical coupling by dipolar filed [2,3].

In magnonics the terms surface states (waves) or edge states (modes) have usually specific meaning. The first one refers usually to the magnetostatic waves in planar system propagating perpendicularly to the in-plain applied magnetic field – i.e. for the so-called Damon-Eshbach (DE) configuration [13, 14]. These waves are localized on the top or at the bottom face of the planar structure, depending on the direction of their wave vector. Magnetostatic surface waves decay exponentially inside the magnetic layer with the rate inversely proportional to their wave length. On the other hand, the magnonic edge modes are the states localized inside the wells of demagnetizing field, located close to the surface of magnetic material in the vicinity of the interface between two magnetic materials. The wells appear inside the material of higher saturation magnetization when the static magnetization has non-zero component normal to the surface/interface and can confine the SWs of the frequencies below the ferromagnetic resonance frequency of a bulk material [14, 15].

In this paper we are going to investigate the surface states in their original meaning known from solid state physics. The surface states in solid state physics were considered initially in electronic systems as a defect states resulting from the termination of infinite periodic structure of the ideal crystal [16, 17, 18,19]. The solution of Schrodinger equation for electronic waves in infinite periodic structures (crystals) have a form of Bloch waves [20]. In the energy ranges corresponding to the energy gaps, the wave vector is complex and solutions expand and decay exponentially for opposite direction of wave vector. Such solutions are rejected because they cannot be normalized in infinite periodic system. However, for the system limited by surfaces, we can accept the solution which decay exponentially inside crystal and match them to the solutions in vacuum, which also decay exponentially for the energies lower than the reference potential in vacuum determined by the work function. Such states bounded at the surfaces of periodic medium (e.g. crystal lattice) are called surface states. This theory of wave excitation in periodic systems can by generalized for any kind of physical system that can be described by the set of linear differential equations with spatially dependent and periodic coefficients [21]. Therefore, in linear regime, we can describe the wave excitation (electromagnetic waves, elastic waves, SWs) in artificial crystals (photonic crystals [22], phononic crystals [23], magnonic crystals [24,25,26], formed by periodic modulation of (electric, elastic, magnetic) material parameters, in the form of Bloch waves. In these systems we can also observe the surface states localized on the boundary between the periodic and homogeneous medium. These states have the frequencies in the ranges of the gaps forbidden for Bloch waves.

The interesting problem is the formulation of the conditions for existence of surface states. We should specify both the bulk parameters (describing the periodic structure and determining the dispersion relation for bulk modes) and surface parameters (describing the way how the surface is introduced in the system) to find the conditions in which surface states appear in the system. This problem was studied in depth for electronic systems, both for simple models[1] and for electronic superlattices [27,28,29]. Such studies were also conducted for photonic [30,31,32] and phononic systems [33]. The very interesting approach to the problem of existence of surface states was presented in the works of K. Artmann [34] and later – J. Zak [35] where two different mechanisms of induction of surface states were proposed. Both authors considered the model of electron scattering in periodic potential of atomic lattice with symmetric wells. They distinguished between Shockley surface states existing in the crystal when the external potential wells are not (or are only slightly) deformed in reference to the potential wells in the bulk region of crystal and Tamm surface states induced by the deformation of external potential wells. For Shockley surface states we can introduce symmetry related criteria [35]. These criteria allow to point out the (energy) gaps allowed and forbidden for Shockley surface states (for given values of bulk parameters). By the deformation of the external wells (described by the surface parameters) the Tamm states can by introduced in the gaps forbidden for Shockley states.

The investigations of surfaces states in magnonics are difficult because of the presence of two general mechanisms of localization [36,37,38,39]: (i) strong inhomogeneity of static demagnetizing field on interfaces of the nanostructure – evoking the localization of edge modes and (ii) time reversal symmetry breaking due to magnetic dipolar interactions for partially confined geometries (planes, wires) – inducing the localization of magnetostatic surface modes/waves. These effects can be present also in the magnonic systems without periodicity. By reference to electronic and photonic systems, we consider as magnonic surface states only the states of frequencies in the ranges of frequency gaps forbidden for Bloch modes, localized on the surfaces braking the periodic structure.

Let's consider the magnonic crystal (MC) for which the inhomogeneous demagnetizing field on interfaces can lead to strong localization of SWs. These edge modes confined in periodically distributed wells of demagnetizing field will form almost flat magnonic band(s) of the frequency shifted, in the spectrum, outside the bands of propagating bulk modes [15]. However, such modes cannot be regarded as surface modes because they are not localized on the surface of MC with decaying amplitude in the bulk region of the structure. On the other hand, we can consider (Tamm) surface states induced by demagnetizing field, produced by termination of periodic structures for which the landscape of demagnetizing field is substantially different in the surface regions of the structure. In our studies we will not investigate this case. For simplicity we will consider the geometry in which the static demagnetizing field is absent and perturbation of the surface regions (necessary for Tamm states) will be introduced by structural changes.
For MC, for which the external field is tangential to the infinitely extended surfaces terminating the periodicity [40], the localization resulting from the presence of periodicity and the localization characteristic for the magnetostatic surface modes may coexist with each other (be intermixed), which aggravate the study of surface states. In our studies we avoid this ambiguity by a) considering the planar structure of finite thickness and b) by neglecting the SW component tangential to the surfaces terminating the periodicity. This component (of nonzero wave number) could be responsible for localization of magnetostatic surface modes on the edges of the first and the last stripes regardless on the periodicity of the system, in the case of the structure of large thickness (when the area of side faces of edge stripes is large).

In this work we plan to adapt the classification of electronic surface states (making a distinction between Shockley and Tamm states) to magnonic surface states [41]. We are going to consider one-dimensional planar MCs, being the finite sequences of two different kind of stripes. The SW dynamics in these systems will be investigated in saturation state where external magnetic field is applied along the stripes. This magnetic configuration allows to cancel static demagnetizing field and to avoid inducing edge modes on surfaces and interfaces. For the considered structures the localization resulting from periodicity (which is essential for appearance of Shockley and Tamm states) and the localization occurring due to time reversal symmetry breaking (all modes propagating from stripe to stripe are DE modes) takes place in orthogonal direction. Shockley and Tamm states have significant amplitude in initial (or/and final) periods of the structure, whereas the DE modes are localized in out-of-plane direction.

The first considered by us system is a direct counterpart of electronic system – magnetic layer with periodically modulated (from stripe to stripe) in-plane anisotropy in exchange dominated regime. The latter one is more general system composed of stripes differing in magnetic material parameters (in magnetization saturation and exchange length) with both exchange and dipolar interactions taken into account. This study will be the significant extension of the research presented in Ref.42 where we investigated two semi-infinite, one-dimensional MCs composed of two kinds of magnetic layers. The semi-infinite MCs were coupled by wide magnetic layer of different kind. In the work [42] we haven't been force to face the problem of boundary conditions on surfaces (interfaces between magnetic material and air) which result from the strength of SW pinning/releasing. We are going to investigate this issue in the presented paper by considering the finite MCs [43] with assumed strength of the surface anisotropy (the first system) and MCs with dipolar pinning (the latter system).

The manuscript is organized as follows. The next section after introduction: 'The model' describes the geometry of investigated structures and their material parameters. We present also the theoretical model which we use to calculate the frequencies and the spatial profiles of SW eigenmodes. In the further section: 'The surface states in anisotropy modulated MC we compare the condition of existence of surface states in electronic system and its magnonic counterpart. Then in the section 'The surface states in bicomponent MC we discuss the magnonic system with dipolar interactions taken into account and spatial modulation of the saturation magnetization and exchange length, which cannot be directly compared to electronic system. In both studies we pay special attention to the role of the strength of surface pinning of magnetization (determined by the surface anisotropy or dipolar interactions) on the condition of existence of the surface states. The systems are investigated numerically with the aid of the plane wave method (PWM) and the finite element method (FEM) (performed in COMSOL Multiphysics package). The main outcomes of our research are summarized in the last section: 'Conclusions'.

## 2. The model

We investigate the surface states in one-dimensional MCs. We consider two systems: (i) the MC with periodic change of the anisotropy field in exchange regime, being the counterpart of electronic system (figure 1(a)), (ii) the MC composed of the Fe and Ni stripes in dipolar regime with exchange interactions included (figure 1(b,c)). The MCs are terminated in symmetry points, i.e. in the middle of the stripes of low or high anisotropy (or the stripes made of Fe or Ni) to implement the symmetry related criteria for existence of Shockley surface states. We investigated also Tamm surface states induced by replacing Ni or Fe stripes in the middle of edge cells by Py wires (green stripes) (figure 1(d,e)) and by the presence of perturbation of the surface areas of the MCs (figure 1(f,g)).

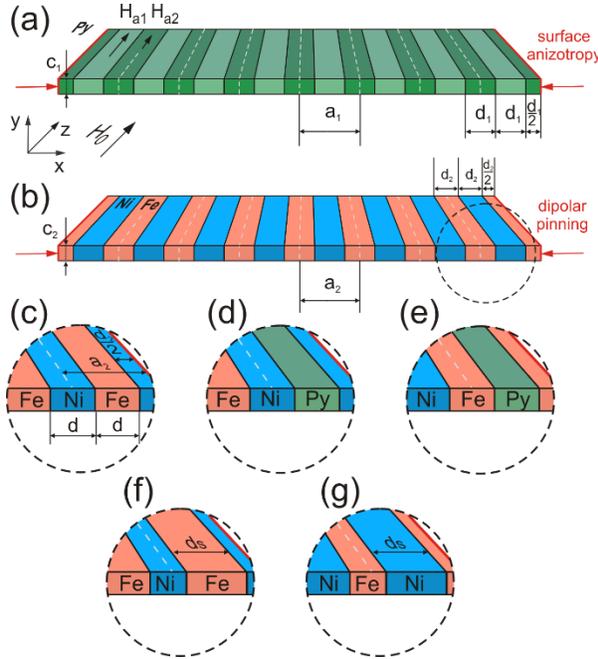

**Figure 1.** Considered MCs. All structures are planar and consist of infinitely long Ni, Fe or Py wires being in the direct contact with each other. The external magnetic field $\mu_0 H_0 = 100$ mT is directed along the wires length and is strong enough to saturate the samples. Every system consist of 8 symmetric unit cells – edges of unit cells are marked by white dashed lines. (a) Anisotropy modulated MCs (in an exchange- regime) consist of homogeneous Py layer with periodically modulated anisotropy magnetic field, taking two alternately changing values: $\mu_0 H_{a1} = 200$ mT (light green stripe) or $\mu_0 H_{a2} = 0$ (dark green stripe). (b,c) Bi-component MCs (in a dipolar-exchange regime) consisting of Ni and Fe wires (blue and red stripes, respectively). We considered two ways of termination of those MCs, both in symmetry points: (b) in the center of Fe stripe, (c) in the middle of Ni stripe. (d,e) MCs originating from (b,c) in which the material in the middle of edge cells (iron – (d), nickel – (e)) was replaced by Py wires (green stripes). (f,g) MCs originating from (b,c) in which the width of material in the middle of edge cells (iron – (f), nickel – (g)) was increased: $d_S > d_2$.

The planar structure presented in figure 1(a) is made of Py layer finite in x-direction (800 nm width) and in y-direction ($c_1$=15nm thickness) but infinite in z-direction. We assume that the static magnetic configuration of considered structure is saturated by the external magnetic field $\mu_0 H_0 = 100$ mT applied in the z-axis. We then introduce the periodical modulation of anisotropy field along the x-axis [44,45,46]. The anisotropy field has got the same direction as the external magnetic field – it is directed along the z-axis. By this we have introduced 8 symmetric unit cells, each of $a_1 = 100$ nm width and $c_1 = 15$ nm thicknesses, characterized by periodically modulated magnetic anisotropy field $\mathbf{H}_{a1} = 200$mT, p or $\mathbf{H}_{a2} = 0$, periodically changing every $a_1/2 = 50$ nm width (along x-axis). The material parameters of Py we used are: saturation magnetization $M_{Py} = 0.7 \times 10^6 \frac{\text{A}}{\text{m}}$, exchange constant $A_{Py} = 1.1 \times 10^{-11} \frac{\text{J}}{\text{m}}$, gyromagnetic ratio $\gamma = 176 \frac{\text{GHz}}{\text{T}}$.

The planar structures (as before – finite in x- and y-direction, infinite in z-direction) showed in figure 1(b-g) are composed of ferromagnetic wires arranged in periodic sequences. The infinitely long wires of the same dimensions are made of Ni, Fe and Py. We investigate MCs consisting of 8 symmetric unit cells made of wires of $d_2 = 250$ nm widths (along the x-axis), $c_2 = 30$ nm thicknesses (along the y-axis) and infinite length along the z-axis. The wires are in the direct contact which ensures the exchange coupling between the successive wires. We have also considered dipolar interaction between those wires. We assume that the static magnetic configuration of considered structures is saturated by the external magnetic field equal to $\mu_0 H_0 = 100$ mT and applied in the direction of the wires axis. The material parameters of Ni, Fe and Py are: saturation magnetizations $M_{Ni} = 0.484 \times 10^6 \frac{\text{A}}{\text{m}}$, $M_{Fe} = 1.752 \times 10^6 \frac{\text{A}}{\text{m}}$, $M_{Py} = 0.86 \times 10^6 \frac{\text{A}}{\text{m}}$, exchange constants: $A_{Ni} = 0.85869 \times 10^{-11} \frac{\text{J}}{\text{m}}$, $A_{Fe} = 2.0992 \times 10^{-11} \frac{\text{J}}{\text{m}}$, $A_{Py} = 1.299774 \times 10^{-11} \frac{\text{J}}{\text{m}}$. We used slightly different value of material parameters for Py in MC with dipolar interaction included to observe the coexistence of Tamm and Shockley states (see figure 3(e) and discussion below). The gyromagnetic ratio $\gamma = 176 \frac{\text{GHz}}{\text{T}}$ is assumed the same for all materials.

To describe magnetization dynamics we solve the Landau-Lifshitz equation (LLE) which is the equation of motion for the magnetization vector **M**:

$$\frac{d\mathbf{M}}{dt} = \gamma \mu_0 \left[ \mathbf{M} \times \mathbf{H}_{eff} + \frac{\alpha}{M_S} \mathbf{M} \times (\mathbf{M} \times \mathbf{H}_{eff}) \right], \tag{1}$$

where: $\alpha$ – is damping coefficient $\mu_0$– is permeability of vacuum, $M_S$ is saturation magnetization and $\mathbf{H}_{\text{eff}}$ – is effective magnetic field. The first term in LLE describes precessional motion of the magnetization around the direction of the effective magnetic field and the second term enrich that precession with damping. We neglect damping in these calculations, putting $\alpha = 0$. The effective magnetic field in general can consist of many terms, but in this paper we will consider: external magnetic field $\mathbf{H_0}$, nonuniform exchange field $\mathbf{H}_{\text{ex}}$, dipolar field $\mathbf{H}_{\text{d}}$ and anisotropy field $\mathbf{H}_{\text{a}}$: $\mathbf{H}_{\text{eff}} = \mathbf{H_0} + \mathbf{H}_{\text{ex}} + \mathbf{H}_{\text{d}} + \mathbf{H}_{\text{a}}$.

We consider the linear regime of magnetization dynamics where we could clearly discuss the SW motion on the background of the static magnetic configuration in saturation state and investigate the SW eigenmodes in the system characterized by harmonic dynamics in time: $e^{i\omega t}$, where $\omega$ is the angular eigenfrequency. The LLE (1) for the magnetization vector **M** can be linearized in the form of set of two differential equations for complex amplitudes of dynamical components of magnetization $m_x$ and $m_y$ [47]:

$$i\frac{\omega}{\gamma\mu_0} m_x(\mathbf{r}) = (H_0 + H_a(\mathbf{r})) m_y(\mathbf{r}) - \frac{2}{\mu_0} \nabla \cdot \frac{A(\mathbf{r})}{M_S(\mathbf{r})} \nabla m_y(\mathbf{r}) + M_S(\mathbf{r}) \frac{\partial}{\partial y} \varphi(\mathbf{r}), \tag{2}$$

$$i\frac{\omega}{\gamma\mu_0} m_y(\mathbf{r}) = -(H_0 + H_a(\mathbf{r})) m_x(\mathbf{r}) + \frac{2}{\mu_0} \nabla \cdot \frac{A(\mathbf{r})}{M_S(\mathbf{r})} \nabla m_x(\mathbf{r}) - M_S(\mathbf{r}) \frac{\partial}{\partial x} \varphi(\mathbf{r}). \tag{3}$$

The anisotropy field and homogeneous external field has only z-component: $\mathbf{H}_a = [0, 0, H_a]$, $\mathbf{H}_0 = [0, 0, H_0]$. Exchange field for the system in DE geometry in saturated state has only dynamical components:

$\mathbf{H}_{ex}(\mathbf{r},t) = [h_{ex,x}(\mathbf{r})e^{i\omega t}, h_{ex,y}(\mathbf{r})e^{i\omega t}, 0]$. The relation between the dynamical exchange field and the amplitudes of dynamical magnetization $\mathbf{m}(\mathbf{r},t) = \mathbf{m}(\mathbf{r})e^{i\omega t} = [m_x(\mathbf{r})e^{i\omega t}, m_y(\mathbf{r})e^{i\omega t}, 0]$ can be expressed in linear approximation as [48]:

$$\mathbf{H}_{ex}(\mathbf{r},t) = \frac{2}{\mu_0 M_S(\mathbf{r})} \nabla \cdot \left(\frac{A(\mathbf{r})}{M_S(\mathbf{r})}\right) \nabla \mathbf{m}(\mathbf{r}) e^{i\omega t}, \qquad (4)$$

The second terms on the right hand side of Eqs. (2) and (3) have exchange origin and result directly from the Eq.(4). Our calculations are based on FEM where the model of continuous medium is investigated. The exchange interaction between the successive stripes is included by appropriate boundary condition at the Ni/Fe interfaces. We use the natural boundary conditions resulting from the exchange operator (4) implemented in the linearized LLE equations (2-3). As was pointed out in [49] these boundary conditions are: (i) continuity of dynamical components of magnetization and (ii) continuity of first derivatives of the dynamical components of magnetization multiplied by factors: $A/M_S$.

For considered geometry the static components of the demagnetizing field are equal to zero, nonzero are only *x*- and *y*-components of the dynamical dipolar field. Using the magnetostatic approximation, the demagnetizing field can be expressed as a gradient of the scalar magnetostatic potential:

$$\mathbf{H}_d(\mathbf{r},t) = [h_{d,x}(\mathbf{r})e^{i\omega t}, h_{d,y}(\mathbf{r})e^{i\omega t}, 0] = -\nabla \varphi(\mathbf{r})e^{i\omega t}. \qquad (5)$$

With the aid of the Gauss equation, we obtain the following equation which relates magnetization and magnetostatic potential:

$$\nabla^2 \varphi(\mathbf{r}) - \frac{\partial m_x(\mathbf{r})}{\partial x} - \frac{\partial m_y(\mathbf{r})}{\partial y} = 0. \qquad (6)$$

The Eqs. (5, 6) can be used to find dynamic components of demagnetizing field implemented already in equations (2-3), i.e., last terms on the right hand side of these equations.
We solve the linearized LLE in the form of the eigenvalue problem (2-3) by the use of FEM with the aid of COMSOL Multiphysics software.

## 3. The surface states in anisotropy modulated magnonic crystal

The MC in an exchange regime is the direct counterpart of electronic system where the saturation magnetization $M_S$ is spatially homogeneous. For one-dimensional system of this type, the linearized LLE (2,3) can be written in the form of Schrodinger equation, using effective mass approximation [42]:

$$-\frac{d}{dx}\lambda_{ex}^2(x)\frac{d}{dx}m_z(x) + v(x)m_z(x) = \Omega\, m_z(x), \qquad (7)$$

where

$$v(x) = \frac{H_0 + H_A(x)}{M_S}, \quad \Omega = \omega \frac{1}{\gamma \mu_0 M_S} \qquad (8)$$

play the role of fictitious potential and energy by comparison to the electronic system. The inverse of the squared exchange length $1/\lambda_{ex}^2$ is the counterpart of effective mass. If an exchange length is also spatially constant then the Eq.7 is mathematically equivalent to the one-dimensional stationary Schrodinger equation. The spatial changes of fictitious electrostatic potential $v(x)$ are then expressed by an anisotropy field $H_a(x)$. It is worth noting that for purely exchange waves the precession is circular and the complex components of dynamical magnetization are related by a simple relation: $m_y(x) = \pm i m_x(x)$, which means that their amplitudes are the same but their harmonic oscillations $e^{i\omega t}$ are shifted by the temporal phase $\pi/2$.

Due to the mathematical form of Eq.(7), the Bloch functions (spin wave eigenmodes) and frequency spectrum of considered MC share properties of the electronic Bloch functions (electronic eigenmodes) and energy spectrum for 1D potential. Therefore, we can almost straightforward apply the formalism describing the condition of existence of electronic surface states, proposed by J.Zak [35,29] to the magnonic surface states in 1D MCs with modulated anisotropy field in exchange regime. To find the electronic eigenmodes in (semi-)finite 1D crystal we can match the logarithmic derivatives at the surface to the solutions in the vacuum and at the crystal. The logarithmic derivative (of electronic wave function) in vacuum $\rho_V$ has constant sign – negative (positive) on the left(right) surface. When the crystal is terminated in the symmetry point of the periodic potential (in the middle of barrier or well) then the logarithmic derivative of Bloch function $\rho_B$ is real and monotonic function of energy in every energy gap, reaching zero and pole on opposite edges of each gap. The sign of $\rho_B$ is then constant for a given gap – the sign of $\rho_B$ is also (similarly like for $\rho_V$) flipped after swapping between left and right surface. Therefore, by matching logarithmic derivatives $\rho_B$ and $\rho_V$, and then comparing their signs, we can point out the gaps forbidden or allowed for the surface states. We can also prove that for semi-infinite crystal terminated in symmetry point only one surface state can appear in one gap. In order to use this formalism for an exchange SWs in MCs we have to impose the boundary conditions on the surface. SWs cannot propagate in nonmagnetic medium (e.g. vacuum) and it is not possible to extend there the solution of Landau-Lifshitz equation. Hence, instead of matching $\rho_B$ and $\rho_V$, we have to compare the $\rho_B$ to the ratio of dynamical magnetization and its spatial derivative on the surface, resulting from the pinning of (exchange) SWs. The arbitrary strength of pinning can be imposed by application of the so-called Rado-Weertman boundary conditions [7]:

$$m_\alpha \mp p \frac{dm_\alpha}{dx} = 0, \qquad (9)$$

where the $\mp$ sign in (9) refers to the left and right surface and index $\alpha = x, y$ denotes different components of the dynamical magnetization.

From Eq.(9), we can notice that inverse of pinning parameter $1/p$ is nothing else than the logarithmic derivative of the dynamical magnetization, and it can be matched directly to the logarithmic derivative of the Bloch function.

The strength of pinning depends on the state of surface and is determined by the surface magnetocrystalline anisotropy. The pinning parameter is then defined by the ratio of surface anisotropy $K_S$ to an exchange stiffness constant $A = \mu_0 M_S^2 \lambda_{ex}^2 / 2$. It means that the value of logarithmic derivative on the surface is fixed and does not depend on the frequency. This behavior is different for electronic surface states where logarithmic derivative in vacuum decreases with the energy $\rho_V \propto \sqrt{(V_V - E)}$. It means that the pinning of an electronic surface state is increasing (with increasing energy), reaching the limit of an ideal pinning for $E = V_V$ (i.e. the maximum energy for which the electronic surface state can be bind at the surface).

We investigated the planar MCs presented in figure 1(a) with spatially modulated anisotropy field. In our studies we solve numerically the general form of linearized Landau-Lifshitz equation (2,3) where we exclude dipolar interaction and assume the homogeneity of saturation magnetization $M_S$ and exchange stiffness constant $A$. We consider planar system where the anisotropy field varies abruptly each 50 nm distance in the x-direction, taking two values: $\mathbf{H}_{a1} = 200$ mT (presence of anisotropy present) or $\mathbf{H}_{a2} = 0$ (presence of anisotropy field) (see figure 1a). We assumed the large value of ratio of lattice constant $a$ (being the doubled width of the stripe) to the thickness $c$ of the structure. This allowed us to avoid the SW quantization along the thickness in the low frequency range. This assumption, being not so far from reality, ensure that the magnonic states, originating from a few lowest magnonic bands, will have practically uniform SW profiles along the thickness. The spectrum of these states will be practically the same as a spectrum of layered system (i.e. systems without confinement in x- and y- direction). To observe the surface states we have limited also our system in the direction of periodicity (x-direction) and took eight complete and symmetric units cells. At the side faces (i.e. at the external edges of the first and the last cell) we applied the Rado-Weertman boundary conditions [7], which allowed us to impose the different strength of magnetization pinning. In calculations, we assumed that the external magnetic field and anisotropy field are oriented along the stipes. However, this particular orientation is not important for exchange SWs. For the case when z-component of a wave vector (component along the stripes) is equal to zero, the considered system can be treated as one-dimensional and can be discussed using the simplified Eqs.(7,8). On the other hand, the geometry and an external field orientation for the considered system (see figure 1(a)) is the same as for the MCs operating in dipolar regime, which we will discuss later (see figure 1(b)). Therefore, the system presented in figure 1(a), in which only the exchange interactions are included, can be treated as a reference one.

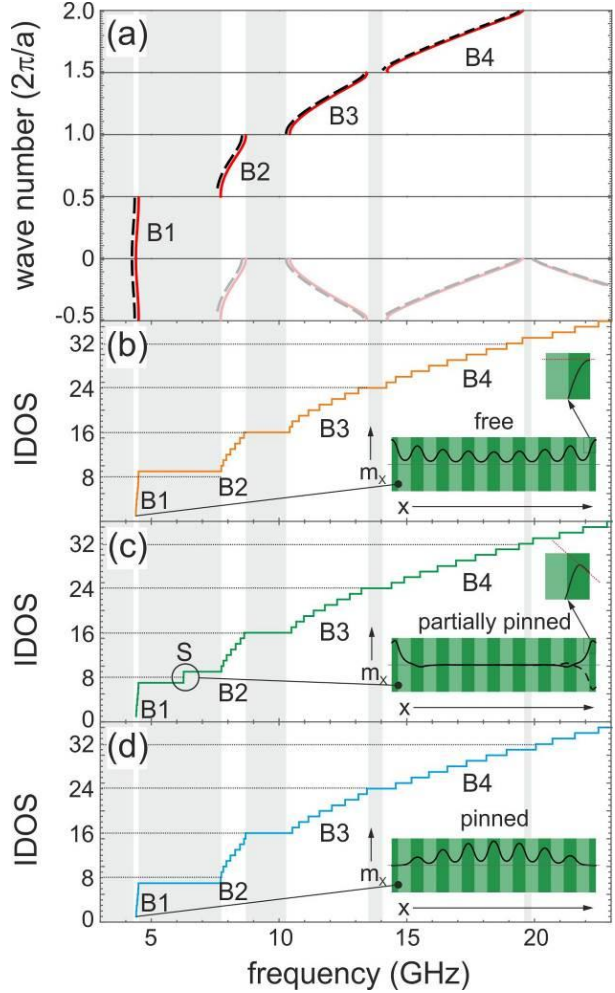

**Figure 2**. The spectra of MC with periodically modulated anisotropy field (see figure1(a)) in an exchange regime, being the direct counterpart of electronic system. The dispersion relation for infinite MC (a) was calculated with the aid of plane wave method (black dashed lines) and finite element method (red lined) to cross-check both methods and to determine the position of frequency bands (B1-B4) and gaps (gray areas) for finite structures (b)-(d). For finite systems, composed of eight symmetric cells, assuming different boundary conditions (i.e. different strength of spin wave pinning taken for the external edges of the first and the last cell, $p$ approaching to $\infty$, and equal 2.82 nm$^{-1}$ and 0 in (b), (c) and (d), respectively) we calculated (using finite element method) the IDOS. We found two (almost degenerate) Shockley surface states (marked by S labeled in (c)) only for partial pinning. The profiles of selected spin wave eigenmodes ($m_x$ component of dynamical magnetization) were plotted in the insets.

Figure 2 presents the outcomes of calculations for the MC with modulated anisotropy field for the discussed above structure. We start our calculations with verifying the model (Eqs.2,3) and the geometry (figure 1(a)) implemented in FEM solver. Initially, we investigated the infinite MC to obtain the dispersion relation (figure 2(a)) and to mark the position of frequency gaps (gray areas in figure 2, separated by the bands B1, B2, …, B4). We compare the dispersion relation calculated using FEM (red line) to the result obtained by PWM (black dashed line). In the FEM calculations we assumed the Bloch boundary conditions linking the edges of one unit cell, while the PWM computations are naturally designed for infinite periodic structures. We achieved good agreement between both methods. In figure 2(b,c,d) we show the spectra of eigenfrequencies for finite MC consisting of eight symmetric unit cells terminated in the middle of the stripe with the higher anisotropy. The spectra are presented in the form of integrated density of states (IDOS)[50]. Dependences of IDOS on frequency are set together with unfolded dispersion relation (figure 2(a)) which approaches to the quadratic dependence $f \sim k^2$. Due to this relation, the successive bands (delimited in the domain of wave number $k$ by successive edges and centers of Brillouin zones) have increasing width and are characterized by growing group velocity.

In figure 2(b,c,d) are presented discussed above dependences obtained for the different strength of pinning on the edges of the external cells. We chose three different values of pinning parameters: $p \to \infty$ (figure 2(b)) – corresponding to completely unpinned magnetization, $p = 2.82$ nm$^{-1}$ (figure 2(c)) – for partial pinning and $p = 0$ (figure 2(d)) – for pinned magnetization. For each of those values we calculated the dependence of IDOS on frequency. The plateaus in these dependences correspond to the ranges of frequency gaps. The distinctive step appearing in the gap between the first and the second band for partially pinned magnetization (figure 2(c)) denotes the presence of double degenerated surface state. We plot the profile of dynamical component of magnetization $m_x$ to check the changes of magnetization pinning for different values of pinning parameter (compare e.g. profiles of the lowest mode in the insets of figure 2(b,d) and to verify the surface or bulk character of the modes. We found that the 8$^{th}$ and 9$^{th}$ modes are localized at the surface which means that they are surface modes (see the inset in figure 2(c)). Because of the mirror symmetry of the whole structure we find the even and the odd surface mode (plotted by solid and dashed lines in the inset) with respect to the center of the MC.

These surface modes can be considered as Shockley surface modes because they appear in the structure terminated in its symmetry points (the center of stripe of higher anisotropy field). The mathematical form of differential Eq.(7), identical to the Schrodinger equation, ensures that all properties of electronic Shockley (and Tamm) states will be observed in this kind of magnonic system. We checked that the following effects are observed in MC with spatial modulation of anisotropy field:

- For every frequency inside given magnonic gap, the logarithmic derivative has constant sign, taken at one of two symmetry points of the structure (the centers of stripe of different kind). The sign of logarithmic derivative determines whether the magnonic gap is allowed or forbidden for Shockley states.
- If we change the unit cell by swapping the areas of low and high anisotropy (and shifting the location of surface by $a_1/2$ to the other symmetry point of the structure) then the sign of logarithmic derivative will be reversed only for the gaps opened at the edge of Brillouin zone (the real part of complex wave number equals $\pi/a$). It means in our case, that the Shockley states in the gap between the first and the second band disappear, but they will be induced in the gap between the third and the fourth band.
- Perturbing the geometry of the MC (or changing the value of anisotropy field) close to its surfaces, we can induce the so-called Tamm states.

Let's discuss the impact of the pinning strength on the conditions of existence of surface states in considered system. We can see that for an ideal pinning (figure 2(b)) or complete unpinning (figure 2(d)) of the magnetization on the surface we do not observe Shockley states. It results from the fact that for an ideal pinning (unpinning) of magnetization the logarithmic derivative of Bloch function at symmetry point $\rho_B$ is matched at the surface to the value $1/p \to \pm\infty$ ( $1/p = 0$) ). For such values of logarithmic derivative $\rho_B$ at the edges of the gaps, the surface states will not be induced. We know also [35] that the logarithmic derivative $\rho_B$ (independently on its sign in given gap) is an increasing function of the frequency in each gap This property implicates that the decrease of pining parameter from $\infty$ (unpinned magnetization) to 0 (pinned magnetization) will shift the position of Shockley state from bottom of the gap to its top, which is observed in figure 2(b-d). It is also worth to notice that by changing the sign of surface anisotropy constant we can reverse the sign of pinning parameter. We do not have such freedom in the case of electronic states where logarithmic derivative in the vacuum must be always positive (negative) on the left (right) surface to ensure the exponential decay of electronic wave function in $\pm\infty$. The change of the sign of pinning parameter swap all gaps allowed for Shockley states to forbidden ones and vice versa. Summarizing this discussion, we can conclude that to observe the Shockley in the given magnonic gap we need the partial pinning of SWs on the surface with the sign of pinning parameter matching the sign of the logarithmic derivative of Bloch function.

## 4. The surface states in bi-component magnonic crystal

After discussing the localization properties of SWs in magnonic systems made of homogeneous magnetic material with spatial modulation of anisotropy field and exchange interactions included (which is a good counterpart of electronic system) we will now investigate the surface states in bi-component MCs in general form in the dipolar-exchange regime. As above, we investigate samples in the DE geometry. This lead to the situation in which there is no static demagnetizing field, whereas the dynamical components of demagnetizing field have got different values for the out-of-plane component and for the in-plane component. In the out-of-plane direction the interfaces with the big contrast of $M_S$ are present, whereas in the in-plane direction the contrast of $M_S$ is smaller. As a result the precession is elliptical. The out-of-plane component of magnetization, in general, is smaller.

    In the figure 3(a) we present the dispersion relation of infinite MC composed of Ni and Fe stripes in the dipolar-exchange regime. We can see that the dispersion relation exhibits the features characteristic for the magnetostatic SWs. The group velocity is constant, non-zero in the vicinity of the Brillouin zone center and it achieves largest values there. With the increase of the wave vector $k$, the group velocity decreases and in the limit of $k \to \infty$ of the pure magnetostatic waves it is equal zero, for dipolar-exchange waves it transforms into a parabolic dispersion for large $k$. It is the reason why each next band is thinner (which is just the opposite case as in the dispersion in the exchange regime). We have calculated them using two methods: the PWM (marked as the dashed black lines) and FEM (marked as the red lines). Two methods were used to cross-check the obtained results and to determine the position of the frequency bands (marked as the B1 – B4) and the bandgaps (marked as the grey areas). We noticed that PWM gives slightly lower position of some magnonic bands. We can attribute this change to the assumption introduced in PWM where ideally uniform distribution of SWs amplitude in out-of-plane direction is considered.

    Firstly we analyze the SW spectra and the SW modes of the finite structure without any surface perturbation – figure 3(b). The structure consists of 8 symmetric unit cells containing the Ni stripe in the center and half of the Fe stripe on each side of the Ni stripe – see the figure 1(b). Bulk modes have greater amplitude in the Ni stripes, i.e. in the material having smaller saturation magnetization (which means lower FMR frequency). In the whole structure we have 8 wells (Ni stripes) in which the spin wave amplitude is concentrating for few lowest bands. There are 8 states for each band because of the coupling between those wells. For this structure, we do not observe any surface states. Note that the SW is fairly well pinned at the edges of external stripes (see the zoomed part of the SW profile in the figure 3(b)). This pinning emerges naturally as a consequence of the occurrence of the dipolar interactions in the finite structure [6] of larger ratio of the width to thickness. It results from the presence of the dynamical components of the dipolar field (it is so-called dipolar pinning).

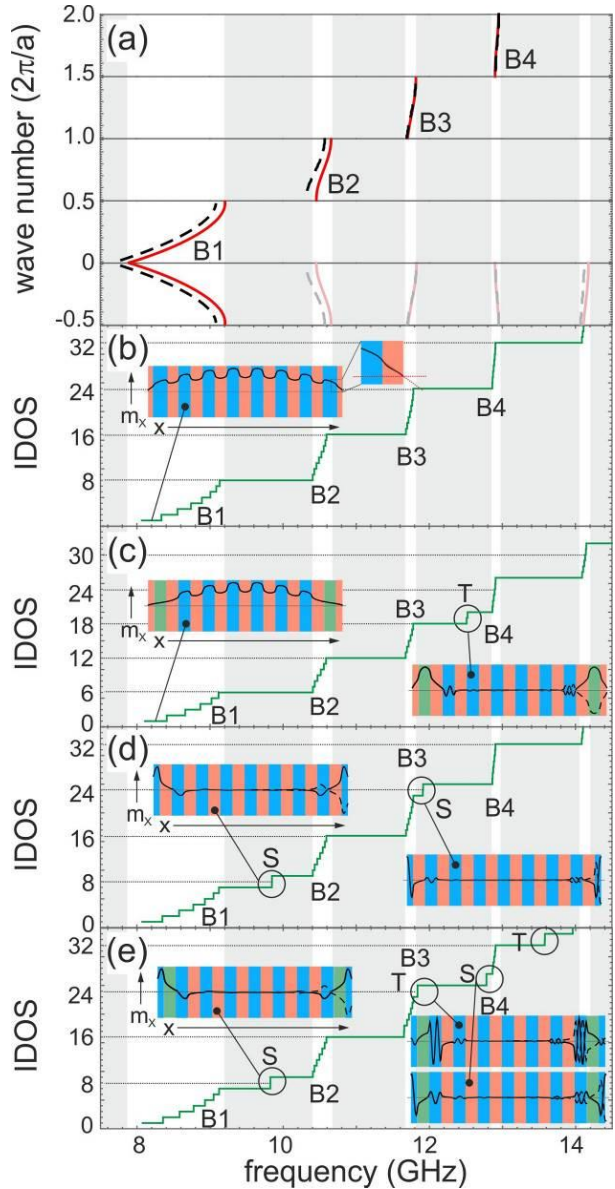

**Figure 3.** The spectra of MC composed of alternately placed Fe and Ni stripes (red and blue bars respectively), considered in dipolar regime – see figure 1(b-g). The dispersion of infinite MC a) calculated by the plane wave method (dashed black lines) and finite element method (red lines) was used to determine frequency bands: B1-B4 and gaps (gray areas). The spin waves in the finite structures: (b)-(e) are partially pinned due to intrinsic mechanism – dipolar interactions. For the structures (b), (d) with external cells unaltered (in reference to bulk ones – see figure 1) we can observe only the Shockley surface states (labeled by 'S'). By replacing the Ni or Fe stripes by Py ones for external cells (dark gray bars) in (c) and (e), respectively, we can induce Tamm surface states (marked by 'T'). The profiles of selected spin wave eigenmodes ($m_x$ component of dynamical magnetization) are plotted in the insets.

Then we have introduced the perturbation of the surface regions of the MC presented in figure 3(b) (see also figure 1(b)) by replacing the Ni stripe in the cells next to the surface by Py stripes, see figure 3(c) (or figure 1(e)). Permalloy, which is the nickel-iron magnetic alloy, have got intermediate material parameters. So it has got higher than Ni, but lower than Fe, FMR frequency. Mentioned replacement results in the lack of concentration of the SWs in bulk modes at the first and at the last unit cells. As a result we obtain 6 wells, in which SWs of low frequencies concentrate. For each lower band B1-B4 we have only 6 modes because of that. Surprisingly, with introducing Py stripes we obtain surface states in this perturbed system. Intuitively, we should expect that introducing Py, as a material of higher than Ni FMR frequency, will not lead to the localization of low frequency SWs in the stripe made of this material. This observation shows that mechanism of induction of surface states of the Tamm type known from electronic theory is general. The crucial factor here is the perturbation of the surface regions in reference to the structure of symmetric cells (see figure 3(b)). The Tamm states, which we found, appear in the third bandgap (between bands B3 and B4). We will show later that by introducing some geometrical changes in the external cells of the MC we can induce the Tamm sates in the other frequency gaps as well.

Thereafter we have investigated the structure (presented in figure 3(d) and figure 1(c)) in which the Fe and Ni stripes are exchanged (in comparison to the figure 3(b)) (or figure 1(b)), so that the Fe stipe is at the center of the unit cell, having one half of the Ni stripe on each side. In this case MC is terminated by the halves of Ni stripes. Redefining the unit cell is indifferent for infinite MCs. However, such change in finite MCs causes shifts of the surfaces from one symmetry point (of infinite structure) to the other one, i.e. from the center of Fe stripe to the center of Ni stripe. According to the findings of J. Zak [35], the change of the symmetry point at which the surface is located turns the gaps opened at the edge of Brillouin zone from allowed (forbidden) to forbidden (allowed) for Shockley states. It seems that this mechanism works also in magnonics for exchange-dipolar waves. In the structure presented in figure 3(d), we observed the appearance of the Shockley states in the first- (between bands B1 and B2) and the third- (between bands B3 and B4) frequency gaps. These gaps are opened at the edge of Brillouin zone where real component of complex wave number is equal to $\pi/a$ (see figure 3(a)). Note that for the structure with halves of the Fe stripes on the edges of the structure (figure 3(b)), we have not found the surface states. The Shockley states that we found (in structure showed in figure 3(d)) are strongly localized exactly at the surface of the MC, i.e. at the Ni stripes of half width.

We have then perturbed the surface regions of MC from figure 3(d) (or figure 1(c)) by replacing the Fe stripe in the cells nearest to the surface by Py stripes – figure 3(e) (or figure 1(d)). We find additional surface states (of Tamm type) appearing in the other gaps, while Shockley-type surface states (appearing in the non-perturbed MC) are still present in the same gaps for perturbed structure presented in figure 3(e). The distribution of SW amplitude for Shockley states in considered structure (figure 3(e)) is similar to those observed in an unperturbed MC (figure 3(d)). Some of Tamm states are found in the same frequency gap as Shockley states (i.e. the gap between bands B3 and B4) – see the insets of the figure 3(e). This property of surface states is different than in electronic crystals or MCs with only exchange and anisotropy included. For these systems (for fixed values of the structural and material parameters), the Tamm and Shockley surface states cannot exist at the same frequency gaps. Here we can find both kinds of surface states appearing at the same gap. There is also other peculiar feature of magnonic systems in dipolar-exchange regime, where the long range interactions are allowed. For simple electronic systems of one-dimensional periodicity, maximum two surface states can exist in one gap. Here, in the bandgap between B3 and B4 bands, we can find four surface states – two Shockley-type (one symmetric and the second one antisymmetric) and two Tamm-type (differing also in the symmetry with respect to the center of the structure). These two Tamm states however are not localized exactly on the surface but on the first full Ni stripe next to the surface.

The Tamm states can be also induced by introduction of structural perturbation to the system. In this approach we can change continuously the strength of perturbation in the surface region of the structure. It was not possible in the previous study where we play with material parameters replacing Fe or Ni stripes by Py in surface cells. Here, we have investigated the behavior of the surface states while changing the ratio $d_S/d_2$ of the widths of the stripes in the surface cells $d_S$ to the widths of regular bulk stripes $d_2$ in the considered MCs.

Firstly we will study MC with the Fe as a middle stripe and Ni as an edge stripe in the unit cell, thus we start with the structure having Ni on the edges, for which the Shockley surface states are present in the absence of perturbation: $d_S/d_2 = 1$ (see figure 3(d)). For this structure, we found Shockley surface states at the first and the third gap for $d_S/d_2 = 1$. It could be also seen in figure 4(a). Then we have made calculations for different values of the ratio: $d_S/d_2$. We kept constant width of the surface cells (while $d_S$ is increasing, the width of the halves of stripes at the edges of cell are decreasing – compare figure 1(c) and figure 1(f)). We have investigated MCs with the described above ratio going from 1 to 1.4. The Shockley surface states (labelled as 'S') travel up to higher frequencies, reaching the frequencies of the above band.

During this process we also observe another surface states appearing. We identify them as Tamm surface states (labelled as 'T'), because the reason of appearing of these surface states is the perturbation of the surface cells. Tamm states emerge from the bands in the gaps above them, and, as it is the case for the Shockley surface states, they travel up to the next band with increasing ratio: $d_S/d_2$. The increase of the frequency of both kinds of surface states with growing value of the ratio $d_S/d_2$ results from reducing the width of Ni stripes in the surface cells. The surface states are localized mostly in Ni stripes and they become more confined with the increase of $d_S/d_2$. Therefore their frequencies are lifted up for higher values of the ratio $d_S/d_2$.

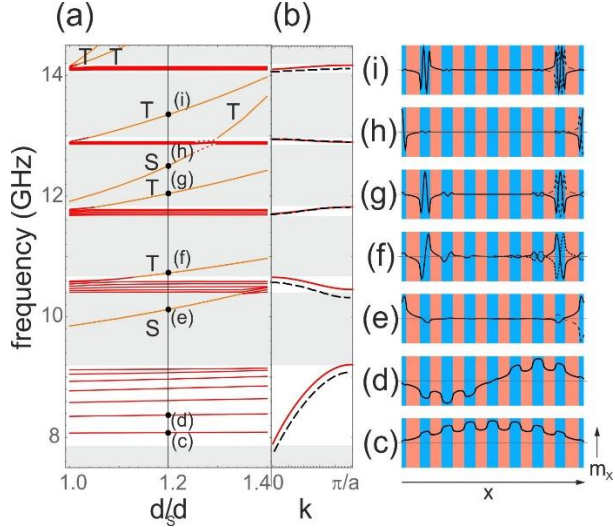

**Figure 4.** (a) Dependence of frequencies of the spin wave eigenmodes in the finite MC (composed of Ni (in blue) and Fe (in red) stripes – see Fig.1(c)) on the perturbation of the external cells (the change of the width of the Fe stripe in external cells $d_S$ with respect to the width of Fe stripes in the bulk cells $d_2$). The variation of the ratio $d_S/d_2$ affects significantly the frequencies of surface states (of Tamm and Shockley type marked be 'T' and 'S') located in energy gaps (gray areas) of infinite MC (the dispersion of infinite system is presented in (b), with red-solid and black-dashed lines obtained from FEM and PWM, respectively). The profiles of the SW eigenmodes ($m_x$ component of dynamical magnetization) were plotted in (c)-(i) for fixed value of the ratio $d_S/d_2 = 1.2$.

We then have plotted the profiles of selected modes for the ratio $d_S/d_2 = 1.2$ in figure 4(c-i). In figure 4(c) and figure 4(d) the first and the second modes (from the lowest band) are presented. In the whole structure, these modes have zero and one node, respectively. Their amplitudes follow the long-scale oscillations characteristic for the metamaterial regime. In figure 4(e) two Shockley states are presented: symmetric mode and antisymmetric one (marked with solid and dashed line, respectively). These modes are localized at the surface cells, mainly at the Ni edge stripe with decreasing amplitude in neighboring Fe stripes and smaller amplitude in the next Ni stripe. We can see the same behavior of the Shockley surface states from the higher bandgap, presented on the figure 4(h). The difference is that Shockley states having higher frequency (figure 4(h)) are more oscillating in space and decay faster inside the MC than the Shockley states from lower gap (figure 4(e)).

The amplitude of the Tamm surface modes is distributed through the structure differently than in the case of the surface states of the Shockley type. We noticed that the SW amplitude, for all of the Tamm states we found, is mostly concentrated under the surface, at the first Ni stripe of the full width (see figure 4(f,g,i)).

It is worth to notice that number of nodal points (or spatial oscillations) of Tamm states increases in successive frequency gaps. The profiles of Tamm states presented in figure 4(f,g,i) have one, two or three nodal points in Ni stripe in the gaps between the bands: B2-B3, B3-B4 and B4-B5, respectively. We can also see that the surface states of Tamm type, similarly like Shockley states, are localized stronger in higher gaps. The strength of localization of surface state depends on the value of imaginary component $k_I$ of the complex wave vector $k = k_R + i\, k_I$, attributed to this state. It is known [1] that the localization is stronger for higher absolute values of $k_I$. The imaginary component of the wave vector has constant sign in the gap and reaches zero on its edges. Moreover, for wider gaps, $k_I$ reaches larger absolute values. Therefore, the strong localization of surface state is observed for the modes existing in the center of wide frequency gaps.

We showed for exchange waves that location of magnonic surface mode inside the gap is related to the strength of the surface pinning. Similar observation for dipolar waves allows us to connect the increase of the localization strength of surface modes (both of Shockley and Tamm type) at higher frequencies to the changes of the dipolar pinning.

In the gaps of higher frequencies (above 11.5 GHz) we can find 4 surface states. In the gap between bands B3-B4, there are two Shockley states (symmetric and antisymmetric) and two Tamm states (symmetric and antisymmetric). For higher gaps all surface states are of the Tamm type. Here, for magnonic system in dipolar-exchange regime, the rule which is in force for the electronic systems, is broken – we found that more than two surface states can appear in one bandgap, and both, Shockley and Tamm surface states, can exist in one bandgap simultaneously.

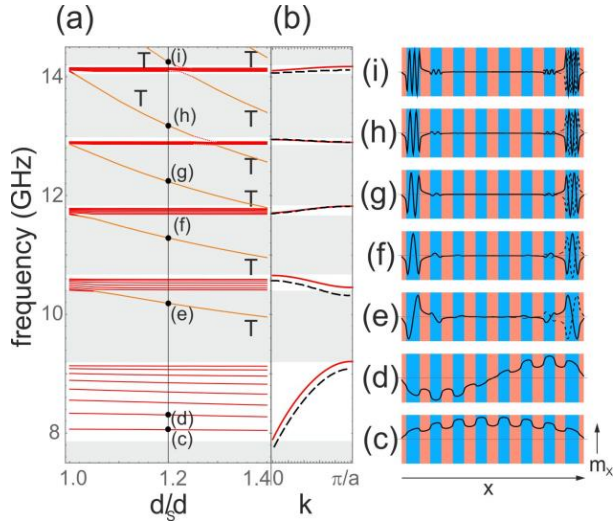

**Figure 5.** (a) Dependence of frequencies of the SW eigenmodes in the finite MC (composed of Ni and Fe stripes – see Fig.1(b)) on the perturbation of external cells (the change of the width of the Ni stripe in external cells $d_S$ in respect to the width of Fe stripes in the bulk cells $d_2$). Note the Ni and Fe stripes are swapped in reference to the structures presented in Fig.4 The variation of the ratio $d_S/d_2$ affects significantly the frequencies of surface states (of Tamm and Shockley type marked be 'T' and 'S') located in energy gaps (gray areas) of infinite MC (the dispersion of infinite system is presented in (b)). The profiles of the SW eigenmodes ($m_x$ component of dynamical magnetization) were plotted (c)-(i) for fixed value of the ratio $d_S/d_2 = 1.2$.

Now we will investigate the behavior of the surface states while changing the ratio $d_S/d_2$ for the finite MC arising due to termination at the second magnetic material (Fe) – see figure 1(b). For this MC there are no surface states in the absence of surface perturbation $d_S/d_2 = 1$. It is interesting to know if it is possible to generate the surface states at this structure simply by changing (increasing) the width of the Ni stripe in the surface cells. It occurs that increasing the width of the middle (Ni) stripe at the surface cell, we can induce the Tamm surface states. For this case, they appear below the bands and will be shifted, with increasing $d_S/d_2$ ratio, to the lower frequencies, and finally reach the region of the lower bands. We can explain this effect in similar manner as for the system discussed in figure 4. The increase of the ratio $d_S/d_2$ extends the width of Ni stripe in which the surface modes are mostly concentrated (see figure 5(e)-(i)). Extending the size of this confinement area results in lowering the frequency of the modes.

The changes of spectrum of MC for different values of ratio $d_S/d_2$ going from 1 to 1.4 are similar as in the previous case (see figure 4), the surface states have got different number of nodal points in successive gaps. However, the lowest Tamm states we have found (of the frequency in the gap between bands: B1-B2), have already one nodal point in the extended Ni stripe. The number of nodal points in mentioned Ni stripe increases systematically for further Tamm states (of the frequencies from successive gaps). In the considered range of structural changes we found for higher gaps (above 11.5 GHz) more than one pair of Tamm states in one gap, which breaks the rule being in force for the electronic systems.

## 5. Conclusions
In an exchange regime, the magnonic system made of homogeneous magnetic material with spatial modulation of the anisotropy field is only a direct counterpart of the electronic systems. The surface states in these systems exhibit the same properties as the electronic surface states in simple 1D models. We can also introduce the clear separation between Shockley and Tamm surface states for these kinds of magnonic systems. The following features of Tamm and Shockley states are sustained in the mentioned sort of magnonic crystals: (i) for the surface located in the symmetry point of the structure we can introduce the symmetry criteria for existence of Shockley surface states, related to the parity of the Bloch function and its derivative in the symmetry point – we can point out in which gaps the Shockley surface states can exist, (ii) shifting the surfaces between two kinds of symmetry points and redefining the (symmetric) unit cell, we can change the gaps opened at the edge of Brillouin zone from allowed for Shockley states to forbidden ones and vice versa (transform the gaps forbidden for surface states into allowed ones), (iii) Tamm and Shockley states cannot exist in the same frequency gaps (for the same values of structural and material parameters), (iv) in one frequency gap can exist maximum two surfaces states. For the magnonic crystals in general form, i.e. with spatial changes of the exchange length and saturation magnetization in dipolar regime (with exchange interactions included), we check numerically that the properties (iii) and (iv) don't hold any more. This makes the clear separation of Tamm and Shockley states in magnonics discursive.

The other difference between the magnonic and electronic surface states is that the SW, in contrary to electronic wave, cannot penetrate the (nonmagnetic) region outside of the crystal. Therefore, the boundary conditions on the surface of magnonic crystal result from the strength of the SW pinning. In an exchange regime the strength of the pinning at the surfaces results from the surface anisotropy which is frequency independent material parameter. The pining parameter play the similar role as logarithmic derivative of electronic wave function in the homogeneous medium (vacuum). The difference is that the 'pinning' of the electronic function on the surface, expressed by the logarithmic derivative, depends on the energy of electronic state. In dipolar dominated regime, the long range dynamical dipolar interactions inside of magnonic crystal are decisive for the pinning of the SWs at the magnonic crystal surfaces.

**Acknowledgements**
We would like to thank Prof. M. Krawczyk for discussion and useful remarks. This work was supported by the National Science Centre Poland for Sonata-Bis grant UMO-2012/07/E/ST3/00538 and the EUs Horizon2020 research and innovation programme under the Marie Sklodowska-Curie GA No. 644348 (MagIC).